\documentclass[prl,twocolumn,showpacs,showkeys,floatfix,amsmath,amssymb]{revtex4}

\usepackage{graphicx}
\usepackage{dcolumn}
\usepackage{bm}

\begin{document}


\title{Modeling Dynamics of Information Networks}

\author{Martin Rosvall}
\altaffiliation{Department of Theoretical Physics, Ume{\aa} University}
\author{Kim Sneppen}%
\affiliation{Nordita, Blegdamsvej 17, 2100 Copenhagen {\O}}

\date{\today}

\begin{abstract}
We propose an information-based model for network
dynamics in which imperfect information
leads to networks where the different vertices
have widely different number of edges to other vertices,
and where the topology has hierarchical features.
The possibility to observe scale free networks is
linked to a minimally connected system where hubs remain dynamic.
\end{abstract}

\maketitle

Complex adaptive systems can often be visualized as networks in which each element is represented by a vertex (node), and its
interactions by edges (links) to other vertices.
Network studies have been inspired by the
observation that working networks often have a broad distribution
of edges and possibly even scale free as reported for the Internet
\cite{albert,Faloustos,Broder2000}, and some molecular networks
\cite{Jeong2000}. Further, real world networks often
exhibit non random topological features. This may be modular
\cite{kleinberg,eckmann,eriksen2}, hierarchical
\cite{maslov_sneppen_prl}, or other features \cite{satorras}, that
e.g.\ may help specificity in signaling \cite{maslov_sneppen_science}.

\begin{figure}
\includegraphics[width=0.7\columnwidth]{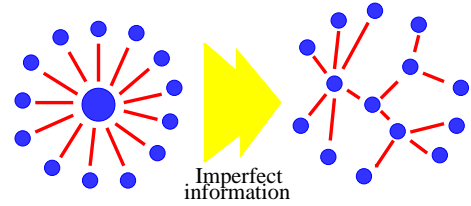}
\vspace{-0.2cm} 
\caption{\label{fig1}%
In a perfect world, a single vertex that can differentiate
all exit edges from each other might distribute all tasks
and information efficiently.
In real world networks, no perfect ''distributor'' exists:
even when every vertex ''tries'' to minimize its distances
to all other vertices, typical vertices tend to connect
through more than one intermediate.
Imperfections destabilize the central hub,
and the vertices in the network
obtain a wide range of vertex degrees.}
\end{figure}

Most networks are the result of a dynamical process.
One hypothesis is preferential growth that predicts scale free networks \cite{barabasi_scale_free,eriksen,newman}. 
The preferential growth is however questionable in many networks,
whereas transmission of information plays a fundamental role in nearly all networks,
including neural networks with synaptic rewiring
\cite{cohen}, molecular networks, and social networks \cite{wasserman},
exemplified by the Internet \cite{Broder2000,Faloustos,huberman,goh}.
In fact, networks may be viewed as the natural embedding of a world with 
a limited information horizon.
Thus, it is interesting to explore a network
topology that is dynamically coupled to information transmission
and formed in an ongoing competition for edges 
between a fixed number of vertices.
We will suggest that a
broad range of vertex degrees could be understood not as an
extension of the narrow distributions of the Erd\H os-R\'{e}nyi
networks \cite{erdos_renyi}, but rather as the result of an intrinsic instability of a
centralized system illustrated in Fig.\ \ref{fig1}.

We consider a dynamic network where each vertex
attempts to optimize its position, given limited information.
A natural quantity to optimize is the participation in the
activities on the network. In economic terms this corresponds
to optimization of trading activity \cite{donangelo}, or to 
maximization of access to a variety of different products. One
activity related measure would be the ''betweenness'' discussed by 
\cite{girvan_newman}. Another measure is vertex--vertex distances,
and accordingly any vertex would attempt to place itself close to all
other vertices. The globally optimized network is then the hub like
structure \cite{valverde}, shown in left panel of Fig.\ \ref{fig1}.
The distances between vertices are minimal and can only
be minimized further by adding additional edges between vertices
on the periphery of the central hub. The addition of such extra
edges is not cost free, as any edge puts a cost to the system. We primarily
consider a dynamics constrained by having the total number of
edges (and vertices) conserved.

In practice each vertex may have only limited information
about the location of other vertices. When changing their
neighbors by moving edges from one vertex to another, they may
make mistakes due to their limited local information.
This will destabilize the optimal topology with a central hub and may lead to a
distributed network as shown in right panel of Fig.\ \ref{fig1}.

To study the interplay between information exchange and dynamical
rewiring of edges in a network, we introduce a simple agent based
model where different agents have different and adjustable
memories in a way reminiscent of the trading model in
\cite{donangelo}. Every agent, named by a number $i=1,2,3 \ldots
n$, is a vertex in a connected network that consists of $N$
vertices and $E$ edges. Agent $i$ has a memory
\begin{displaymath}
M_i=
\left\{ \begin{array}{c}
D_{i}(l) \\
P_{i}(l) \\
\end{array}\right.,\:l=1,2,\ldots,i-1,i+1,\ldots N,
\end{displaymath}
with $N-1$ \emph{distances} $D$ and \emph{pointers} $P$ to the
other agents in the network. The distance $D_{i}(l)$ is agent
$i$'s estimated shortest path length to $l$. The pointer
$P_{i}(l)$ is agent $i$'s nearest neighbor on the estimated
shortest path to $l$. Thus $M_i$ may be seen as a simplified
version of the gateway protocol used by the autonomous systems to
direct transmission of E-mails across the hardwired Internet.
Here, however, the memory will be used to
rewire edges in the network.

\begin{figure}
\includegraphics[width=0.7\columnwidth]{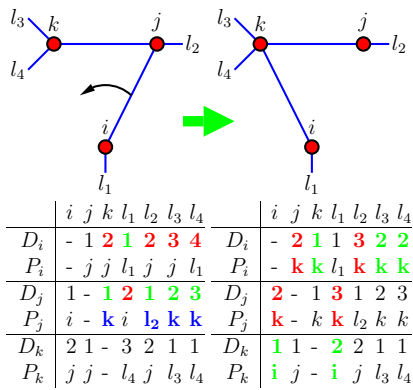}
\vspace{-0.5cm} 
\caption{\label{rewire}%
Dynamics of edge rewiring:
The edge between $i$ and $j$ is rewired to an edge between $i$ and $k$,
if local information predicts that
$k$ provides a shorter path to the random agent $l$ ($l=l_3$ in figure).
The agents' information about the network is subsequently updated as shown by the
shift from lower left to lower right panel.
Notice that the local information not necessarily is correct.
\label{fig2}}
\end{figure}

Initially the network is a hub of the $N-1$ agents connected to a
center agent by $N-1$ edges (as in Fig.\ \ref{fig1}, left) plus $E-N+1$
randomly placed edges on the periphery of the hub.
The basic move, illustrated in Fig.\ \ref{fig2}, consists of a rewiring attempt plus some information exchange 
in a local region of the network.
In detail the move consists of three steps:
\begin{list}{}{}

\item (i) An agent $i$ and one of its neighbors $j$ is chosen
randomly.

\item (ii) An agent $l \ne i,\:j$ is chosen randomly and
if \mbox{$D_{i}(l) >  D_{j}(l)$} then the edge between $i$ and $j$
is rewired to an edge between $i$ and $k=P_j(l)$. If $l$ did not
satisfy the above criteria a new $l$ is randomly chosen. If no
such $l$ exists the rewiring is aborted.

\item (iii)  The information $i$
has lost by disconnecting $j$ is replaced by information from $k$.
Further, there is full exchange of information between
$i$ and $k$: If agent $k$ lists a shorter path to some other agents,
then $i$ adopts this path with a pointer to $k$. Similarly for
$k$, if agent $i$ lists a shorter path then $k$ adopts this path
through $i$.
The information $j$ has lost by disconnecting
$i$ is replaced by forcing agent $j$ to
change all its previous pointers toward $i$ to pointers toward
$k$ and add 1 to the corresponding distances.

\end{list}

Notice that above there is no information transfer between $j$ and
$k$: $j$ does not read any of $k$'s information, $j$ is only using the
information that the rewiring took place. The model defines an update
of both the network (ii) and the information that agents in
the network have about each other's locations (iii). The step (ii)
represents local optimization where agent $i$ rewires from $j$ to
$k$ with a probability given by the fraction of the network which
is estimated to be closer to the center. We stress that only a small part
of the system is informed about a changed geometry and that
decisions on moves may be based on outdated information. When
repeated many times the model leads to a break down of the central
hub into a steady state ensemble of networks with a broad
distribution of vertex degrees.

Fig.\ \ref{fig3}b shows that the degree distribution for vertices in
the network is broad, in fact close to the Zipf law $1/C^2$
reported for some real world networks \cite{Faloustos,Jeong2000},
as well as for the size distributions of industrial companies
\cite{zajdenweber}. However, there is correction to scaling at
intermediate and large vertex degrees. This limitation of the
model can be removed by increasing the information between agents
during the rewiring,
for example by adding information exchange between 
agent $j$ and agent $k$ in Fig.\ \ref{fig2}:  
\begin{figure*}
\includegraphics[width=0.9\textwidth]{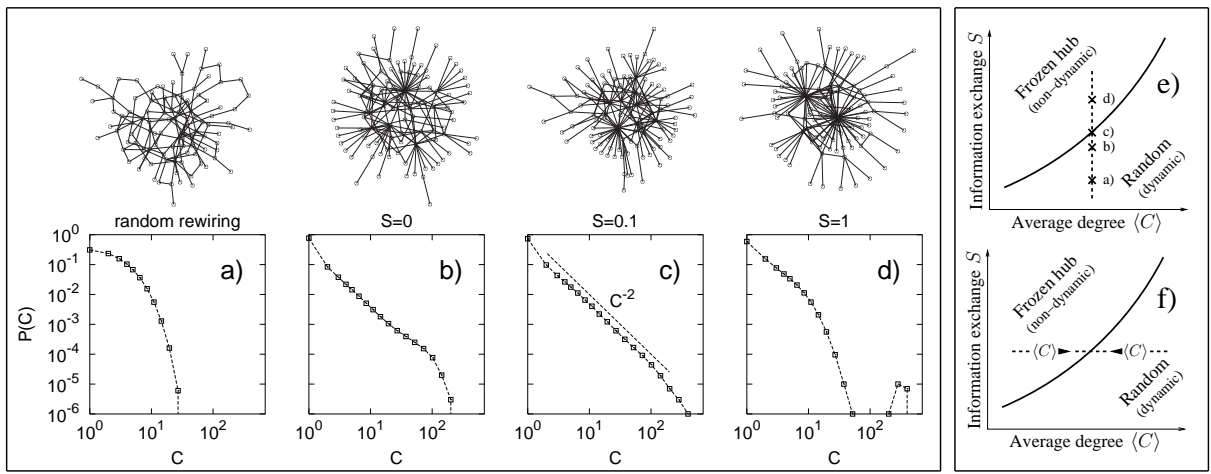}
\vspace{-0.2cm} 
\caption{
\label{fig3}
\emph{Left panel}: Vertex degree distribution of the evolved network with 4
levels of information exchange: no exchange, i.e.\ only rules (i) and (ii) apply in \textbf{a)}, full exchange as in (iii) with exchange of rate $S$ in (iv) with $S=0$, $S=0.1$, and $S=1.0$ in \textbf{b)}--\textbf{d)}. In all lower cases we sample
dynamics of an $N=1000$ vertex system with $E=1500$ edges
($\langle C \rangle =3$). The plots show average of many samples.
The upper graphs show the corresponding networks of size $N=100$.
\emph{Right panel}: Schematic phase diagram illustrating the critical line which separates the dynamic and non-dynamic regime. The information exchange at levels \textbf{a)}--\textbf{d)} from left panel in \textbf{e)} and the variation of $\langle C \rangle$ that drives the network towards the critical line in \textbf{f)}.}
\end{figure*}

\begin{list}{}{}
\item (iv) $j$ considers a fraction $S$ of the information it has
stored with a pointer toward $k$. For this fraction it is checked
whether $k$ lists a shorter path than $j$. For each path where this
is the case, the memory of $k$ is used to update the memory of
$j$.
\end{list}

Notice again that the update in (iv) takes place no matter which
agent had the right data. When $S=0$ the result is as in the simple
model (i-iii), whereas $S=1$ leads to a hub like structure
illustrated with the isolated distribution of highly connected
vertices in Fig.\ \ref{fig3}d. In between there is a critical value
of $S=S_{crit} \sim 0.1$ (for $\langle C \rangle=3$) where one
obtains a scale free distribution of vertex degree (Fig.\ \ref{fig3}c).
$S_{crit}$ depends on the overall edge density
in the system, and increases as the average degree $\langle C
\rangle$ increases. Decreasing $\langle C \rangle$ below $2.9$ even
$S=0$ becomes super critical and the central hub of a big system
($N >> 100$) will never break down. Oppositely, it is remarkable
that an increase in $C$ for fixed $S$ makes it increasingly
difficult to obtain vertices with very high $C$. In any case, at
conditions when one hub dominates the topology, the hub becomes frozen and will never break down. 
Clearly, a scale free degree distribution requires an instability 
and the possibility for vertices to change status dynamically. On
the other hand, when the instability becomes too large, no large
hubs develop and the degree distribution becomes exponential.

\begin{figure}
\includegraphics[width=0.48\textwidth]{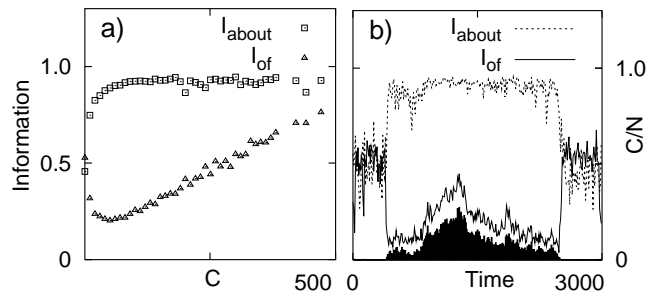}
\vspace{-0.6cm} 
\caption{\label{fig4} {\bf a)} Average information related to agents with
vertex degree $C$ for a simulation with critical information
exchange.
The upper curve is the fraction of agents with correct
information $I_{about}$ about their paths to the specific agent of
degree $C$.
The lower curve similarly refers to the information
$I_{of}$ the agent with degree $C$ has about paths to other
agents. 
{\bf b)} Trajectory for a specific agent with its vertex
degree (dark shaded area), the information the system has about the agent, $I_{about}$, and
the information the agent has about the system, $I_{of}$. Time is counted as number of rewiring
updates per agent. }
\end{figure}

\begin{figure}
\includegraphics[width=0.7\columnwidth]{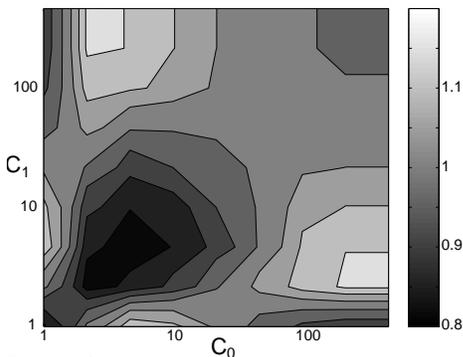}
\vspace{-0.5cm} 
\caption{Correlation profile for an ensemble of model
networks with $\langle C \rangle=3$.
The correlation profile measures the probability for an edge between two
vertices of degree $C_0$ and $C_1$ in units of what it would
be in a properly randomized network. One notices that agents with $C \sim 1$
often connect to agents with $C \sim 5$ that preferentially connect
to agents with high $C$. Thus the network exhibits hierarchical
features. \label{fig5}}
\end{figure}

For simplicity we in Fig.\ \ref{fig4}-\ref{fig5} consider the case of N=1000, E=1500,
and thus $\langle C \rangle=2E/N=3$ with $S=S_{crit}=0.1$. We
stress that the reported results are similar for other values of
$\langle C \rangle$, provided that $S$ is not too far from
$S_{crit}(C)$. E.g. $S_{crit} ( \langle C \rangle =2.5
)=0$ and $S_{crit}(\langle C \rangle = 5 )=0.45$. Also it is
important to stress that the particular choice of rewiring attempt
and information exchange in the above model is somewhat arbitrary.
Therefore we have tested robustness of the obtained results
against a number of variations, including selection of agent $i$
with weight proportional to its degree, aborting step (ii)
after only one attempted $l$, and introducing information exchange
between $i$ and $j$. In all cases we are able to reproduce the
qualitative features of Fig.\ \ref{fig3}-\ref{fig5}.
In particular, a higher overall edge density always requires
a higher information exchange to obtain similar large hubs, as illustrated in Fig.\ \ref{fig3}e. 
For any amount of information exchange a scale free network is obtained for the minimal $\langle C \rangle$ where the hubs remain dynamic (Fig.\ \ref{fig3}f).

Figure \ref{fig4} shows a) the average information
content related to agents of vertex degree $C$ and b) the
temporal development of one particular agent. In both panels,
$I_{of}(i)$ is the fraction of the information $i$ has about
distances and directions to all other agents that is correct.
Information $I_{about}(i)$ is defined as the fraction of other
agents that have correct information about their paths to $i$. The
upper curve in Fig.\ \ref{fig4}a shows that the system
systematically increases the $I_{about}(i)$ as the vertex degree
of $i$ is increased. More surprisingly is the non monotonous
behavior of $I_{of}(i)$:  Agents with intermediate vertex degree
$C$ know the least about the system. They are messed up by false
information about directions, whereas the lowly connected agents
are better informed through their typically higher connected
neighbor.

Figure \ref{fig4}b follows a particular agent through a period of success, 
where it evolves to become one of the major hubs in the
system. The figure shows both the degree of the agent, and the
information related to it. Notice that an initially moderate
increase in degree $C$ at time $\sim 500$ triggers an increase in
$I_{about}$ and a sharp decrease in $I_{of}$. Subsequent increases
in $C$ have little effect on the near perfect information that the
system has about the agent, but a roughly proportional effect on
the quality of the information $I_{of}$. Thus the trajectory of
a particular agent again reflects the ease at which one may
locate anybody in or above the ''middle class'', and the
exclusiveness of having system-wide correct information.

To explore the connectivity pattern between low and high connected
agents, we in Fig.\ \ref{fig5} investigate the correlation profile
of the evolved network \cite{maslov_sneppen_science}. This
quantifies the tendency of agents with different vertex degrees to
connect to each other, by normalizing to a randomized network
where degrees of all vertices are exactly maintained
\cite{maslov_sneppen_prl}. We see that all types of connections
exist, but also that there is a tendency towards hierarchical
organization: Agents with $C \sim 1$ often connect to agents with
degree $C \sim 5$, that preferentially connect to agents with very high $C$. 
This hierarchical pattern is also seen at other values of
$\langle C \rangle $, with decreased amplitude as $\langle C \rangle$ is
increased. Going in the opposite direction, towards decreasing
$\langle C \rangle$, our standard model quickly becomes super critical
even for $S=0$. This can be adjusted by decreasing the
information transfer between $i$ and $k$ in step (iii)
such that this transfer is less than complete.

It is interesting to explore the sociological implications of
the proposed network dynamics, e.g.\ the response to increased information associated to a particular
agent. If we start with an agent of degree $C=1$ and from
this instant keep it perfectly informed about the position of all
other agents, $I_{of}(i)=1$, the result is insignificant.
Similarly, when an agent constantly broadcasts its correct
position to all other agents, that is $I_{about}(i)=1$, the agent
only performs slightly better than average. However, an agent that
allows all its neighbors to update their information by using
his information, very quickly becomes a central hub in
the system. This happens in spite of the fact that his information may
be as bad as that of anybody else. Communication, not correctness,
is the key to success.

Finally we reiterate that the
critical line in Fig.\ \ref{fig3}f corresponds to
the minimal $\langle C\rangle$
where the major hub remains dynamic. 
This suggests a principle
in which the network could self organize 
to become scale free.
This idea is investigated by allowing agents, at a low rate,
to create and destroy edges with probabilities $P_c$ and $1-P_c$,
dependent on the dominance of the major hub.
That is, we set $P_c$ to be an increasing function of the 
dominance of the largest hub, reflecting a 
situation where links are 
created in a persistently centralized system and removed in an 
unstructured system.
For example $P_c = 1- C_2/C_1$, where $C_1$ and $C_2$ 
are the highest and next highest degree in the network,
results in a system that self organizes around the 
critical line as shown in Fig.\ \ref{fig3}f.

The present work suggests a dynamical model where networks with 
both small and large hubs emerge from local
optimization of activity through guesses based on imperfect information.
The frame is formulated in an agent based model,
which is comparable to a sociological setting.
For static snapshots the model predicts a hierarchical organization of
vertices with the highly connected vertices in the center.
This is a plausible feature of business networks and
a quantifiable characteristic of the hardwired Internet
\cite{maslov_sneppen_prl}.
\begin{acknowledgments}
We warmly thank Petter Minnhagen and Ala Trusina for stimulating discussions.
\end{acknowledgments}

\end{document}